\documentclass{amsart}

\usepackage{amsthm}
\usepackage{amssymb}
\usepackage{xspace}

\newcommand{\CC}{\mathbb{C}}
\newcommand{\QQ}{\mathbb{Q}}
\newcommand{\RR}{\mathbb{R}}
\newcommand{\ZZ}{\mathbb{Z}}
\newcommand{\FF}{\mathbb{F}}
\newcommand{\infinity}{\infty}
\renewcommand{\d}{\mathrm{d}}
\newcommand{\e}{\mathrm{e}}
\renewcommand{\i}{\mathrm{i}}
\newcommand{\tr}{\mathop{\mathrm{tr}}}
\newcommand{\ket}[1]{|#1\rangle}

\newcommand{\braket}[2]{\langle#1|#2\rangle}
\newtheorem{example}{Example}

\newtheorem{lemma}{Lemma}
\newtheorem{theorem}{Theorem}
\newtheorem{corollary}{Corollary}
\newcommand{\MUBs}{\textsc{mub}s\xspace}

\usepackage{amsrefs}
\BibSpec{article}{%
    +{}  { \PrintAuthors}                {author}
    +{,} { \textit}                     {title}
    +{.} { }                            {part}
    +{:} { \textit}                     {subtitle}
    +{,} { \PrintContributions}         {contribution}
    +{.} { \PrintPartials}              {partial}
    +{,} { }                            {journal}
    +{,} { \voltext}                    {volume}
    +{,} { \issuetext}                  {number}
    +{,} { pp.~}                        {pages}
    +{}  { \PrintDatePV}                {date}
    +{,} { }                            {status}
    +{,} { \PrintDOI}                   {doi}
    +{,} { \eprint}                     {eprint}
    +{}  { \parenthesize}               {language}
    +{}  { \PrintTranslation}           {translation}
    +{;} { \PrintReprint}               {reprint}
    +{.} { }                            {note}
    +{.} {}                             {transition}
    +{}  {\SentenceSpace \PrintReviews} {review}
}

\begin{document}

\title[MUBs for Quantum States defined over ${p}$-adic numbers]{Mutually unbiased bases for quantum states defined over $\boldsymbol{p}$-adic numbers}
\author{Wim van Dam}
\address{Wim van Dam, Department of Computer Science, Department of Physics, University of California, Santa Barbara, California, 93106, United States of America}
\author{Alexander Russell}
\address{Alexander Russell, Department of Computer Science, University of Connecticut, Storrs, Connecticut, 06269, United States of America}

\begin{abstract}
  We describe sets of mutually unbiased bases (\MUBs) for quantum states defined over the $p$-adic numbers $\QQ_p$, i.e.\ 
the states that can be described as elements of the (rigged) Hilbert space $L^2(\QQ_p)$.  
We find that for every prime $p>2$ there are at least $p+1$ \MUBs, which is in contrast with the situation for quantum states defined   over the real line $\RR$ for which only $3$ \MUBs are known. 
 We comment on the possible reason for the difference regarding \MUBs between these two infinite dimensional Hilbert spaces.
\end{abstract}

\maketitle
% \tableofcontents

\thispagestyle{empty}
% \newpage
\section{Mutually Unbiased Bases}
Let $V=\{v_1,\dots,v_d\}$ and $W=\{w_1,\dots,w_d\}$ be two orthonormal bases for the $d$-dimensional Hilbert space 
$L^2(\ZZ/d\ZZ)\simeq \CC^d$. We say that $V$ and $W$ are \emph{mutually unbiased bases} (\MUBs) if and only if all vectors of
$V$ are a uniform, unbiased superposition in terms of the $W$ vectors, and vice versa. It is straightforward to show 
that this is equivalent with the requirement $|\braket{v_i}{w_j}|=1/\sqrt{d}$ for all $i,j\in\{1,\dots,d\}$. 
A set $\{V_1,...,V_N\}$ of bases is mutually unbiased if and only if each proper pair $V_k, V_\ell$ (with $k\neq \ell$) of 
bases is mutually unbiased. As an example, for the qubit $d=2$ case we have a set of three \MUBs: $V_1=\{\ket{0},\ket{1}\}$, 
$V_2 = \{(\ket{0}+\ket{1})/\sqrt{2},(\ket{0}-\ket{1})/\sqrt{2}\}$, and $V_3=\{(\ket{0}+\i\ket{1})/\sqrt{2},(\ket{0}-\i\ket{1})/\sqrt{2}\}$. 
Mutually unbiased bases play an important role in the problem of optimal quantum state estimation \cite{WoottersFields}, 
quantum cryptography \cite{Cerfetal} and 
the construction of discrete Wigner functions \cite{Cormicketal}. 

A central question in the theory of finite dimensional \MUBs is what the maximal cardinality $N(\CC^d)$ is 
of a set of \MUBs for a given dimension $d$. 
It can be shown that $N(\CC^d) \leq d+1$ and we also know that if $d$ is a power of a prime, $p^r$, 
then this bound can achieved: $N(\CC^{p^r}) = p^r+1$ . (Below we will give an explicit construction of $p^r+1$ \MUBs in $\CC^{p^r}$.) 
For dimensions that are not a prime power, the same question is wide open. For $d=6$ we know that there are $3$ \MUBs and several extensive numerical computations suggest that this is largest possible number of \MUBs in $\CC^6$. However, despite 
significant efforts on this $N(\CC^6)=3?$ question, we still do not have a rigorous proof that excludes the possibility 
that $N(\CC^6)=7$. More generally for $d=p_1^{r_1}\cdots p_k^{r_k}$ it is known that 
$N(\CC^d) \geq 1+\min_i p_i^{r_i}$ (Lemma~3 in \cite{ConstructionsofMUBs}), but to strengthen this bound has proven to be 
difficult, although not impossible for certain very specific cases \cite{WocjanBeth}.  

% Mutually unbiased bases are also related to the construction of sets of 
% \emph{symmetric, informationally-complete} quantum states 
% (\textsc{sic} states) $\{v_1,\dots,v_m\}$ in $\CC^d$, which have the inner product 
% property $|\braket{v_i}{v_j}| = 1/\sqrt{d+1}$ for all 
% $i\neq j$. Zauner's conjecture states that for each finite dimension $d$, there 
% exists $d^2$ \textsc{sic} states in  $\CC^d$. Again, 
% extensive numerical computations support this conjecture, but we seem to be far 
% away from a proof of this elegant conjecture. 

\subsection{The ``$\boldsymbol{(ax^2+bx)}$ construction'' of $(\boldsymbol{p^r+1})$ MUBs in dimension $\boldsymbol{p^r}$} 

It is notable that, although the question of \MUBs 
% and \textsc{sic} states 
is a purely geometric one, the constructions often involve nontrivial results in number theory. 
The following construction of a 
maximal set of $p^r+1$ mutually unbiased bases in $\CC^{p^r}$ is a quintessential example of this \cites{Alltop,ConstructionsofMUBs}.  
\begin{example}
Let $p$ be a prime and $r\in\ZZ^+$ a positive exponent. Using the finite field $\FF_{p^r}$ and its trace 
operation $\tr:\FF_{p^r}\rightarrow \FF_p$ with $\tr:x \mapsto x+x^p+x^{p^2}+\cdots + x^{p^{r-1}}$, we define the following set 
of $p^r$ bases $V_a$ indexed by $a\in\FF_{p^r}$: 
\begin{align}
V_a := \{\ket{v(a,b)}: b\in\FF_{p^r}\}
\mbox{~with~} 
\ket{v(a,b)}  := \frac{1}{\sqrt{p^r}}\sum_{x\in\FF_{p^r}}\e^{2\pi\i \tr(ax^2+bx)/p}\ket{x}. 
\end{align}
Combined with the computational basis $V_{\infinity} := \{\ket{b}:b\in\FF_{p^r}\}$, the set of $p^r+1$ bases
$\{V_a : a\in\FF_{p^r}\cup\{\infinity\}\}$ is mutually unbiased. This fact can be proven using standard results on 
quadratic Gauss sums \cite{GaussSums}.
 That the $V_\infinity$ vectors are mutually unbiased to the $V_a$ vectors follows trivially from the fact that all amplitudes of all 
 $v(a,b)$ have norm $1/\sqrt{p^r}$. To prove that the $V_a$ bases are mutually unbiased one should first observe
 that 
 \begin{align}
 \braket{v(a',b')}{v(a,b)} & = \frac{1}{p^r}\sum_{x\in\FF_{p^r}}\e^{2\pi\i ((a-a')x^2+(b-b')x)}. 
 \end{align} 
 The mutually unbiasedness is then proven by the following result on quadratic Gauss sums over finite fields
 \begin{align}
 \left|\sum_{x\in\FF_{p^r}}\e^{2\pi\i\tr(\alpha x^2+\beta x)}\right| & = 
 \begin{cases}
 \sqrt{p^r} & \mbox{if $\alpha\neq 0$} \\
 0 & \mbox{if $\alpha = 0$ and $\beta \neq 0$} \\
 p^r & \mbox{if $\alpha = 0$ and $\beta = 0$.} 
 \end{cases}
 \end{align}
\end{example}
 Gauss sums $\sum_s \e^{2\pi\i(ax^2+bx)/d}$ over a ring $\ZZ/d\ZZ$ behave `less nicely' when $d$ is not prime, 
 which prevents the above $(ax^2+bx)$ construction to work for arbitrary dimensions $d$. Nevertheless, we
 will use this construction to find \MUBs for infinite dimensional Hilbert spaces. 

\subsection{MUBs for infinite dimensional Hilbert spaces?}\label{sec:L2RMUBs}
It is a non-trivial question to ask if one can define mutually unbiased bases for infinity dimensional Hilbert space 
and, if this is indeed possible, how many \MUBs exist in such spaces. Weigert and Wilkinson\cite{MUBsContinuousVariables} 
did exactly this for quantum variables defined over the real line and they found only 
$3$ \MUBs in this Hilbert space $L^2(\RR)$. The three bases they found were the (generalized) eigenvectors of operators that are a combination of the position operator $\hat{q}$ and the momentum operator $\hat{p}$;  the three operators they used are $\hat{q}$, $\cos(\frac{2\pi}{3})\hat{q}+\sin(\frac{2\pi}{3})\hat{p}$ and $\cos(\frac{2\pi}{3})\hat{q}-\sin(\frac{2\pi}{3})\hat{p}$. 
% The generalized eigenstates of the last two operators $-\frac{1}{2}\hat{q}\pm \frac{\sqrt{3}}{2}\hat{p}$ 
% can be described by the two functions 
% \begin{align}
% v(\pm,b):\RR \rightarrow \CC \mbox{~with~}v(\pm,b):x \mapsto \e^{\pm\i(x^2/\sqrt{12}+bx)}
% \end{align}
% for all $b\in\RR$. The inner product between these two function obeys $|\langle v(-,b'),v(+,b)\rangle| = \sqrt[4]{3}\sqrt{\pi}$. 

\subsection{Our Result}
Partly in response to the $N(L^2(\RR))\geq 3$ result of Weigert and Wilkinson, Blume-Kohout wondered  `Is $\infinity$ prime? or possibly a multiple of $2$?'' \cite{Blume-Kohout}. The idea being that if the infinite
dimension of $L^2(\RR)$ is `prime', we would expect $N(L^2(\RR)) = \infinity + 1$, whereas if the dimension is even, 
then maybe indeed $N(L^2(\RR))=3$, just as we seem to have $N(\CC^6)=3$. Our result here shows that the answer 
to Blume-Kohout's question likely depends on properties of the Hilbert space that go beyond the fact that its 
dimensionality is infinite. We do this by considering quantum states that are defined over the $p$-adic numbers 
$\QQ_p$, i.e.\ we look at \MUBs in the infinite dimensional Hilbert space $L^2(\QQ_p)$. Using the
$(ax^2+bx)$ construction of the previous section and the theory of quadratic Gauss integrals over $\QQ_p$, 
we show that for each prime $p>2$ one has the lower bound $N(L^2(\QQ_p))\geq p+1$. 

For functions in $L^2(\QQ_p)$ we have a well-defined notion of continuity and $\QQ_p$ 
is an infinite, locally compact Abelian group, just like $\RR$. In this sense, $\QQ_p$ defines a continuous variable 
that is different from $\RR$ and one can study the idea
of ``$p$-adic quantum mechanics'', which has been done already by various authors \cites{Kozyrev,QMonpadicfields,Vladimirov}. 
In the next section we will provide the necessary definitions to work with the Hilbert space $L^2(\QQ_p)$. 
After that, in Section~\ref{sec:padicMUBs}, we describe a construction of $p+1$ \MUBs in this space, and prove
their mutually unbiasedness. 
 In the last two sections we address the question why \MUBs for real valued 
 variables seem to behave differently from those defined over $\QQ_p$.

\section{$p$-adic Quantum Mechanics}
\subsection{$\boldsymbol{p}$-adic numbers}
An excellent introduction to $p$-adic numbers is provided by Gouv{\^e}a \cite{Gouvea}; here we will mostly recite the necessary basic definitions.  
Consider the possible norms on the field of rational numbers $\QQ$. Besides the standard $\RR$ norm, we also have the \emph{$p$-adic norm} $|\cdot|_p$ for each prime integer $p$ as a possibility, which is defined as follows. 
For $x\in \QQ\setminus\{0\}$, write $x = p^v (a/b)$ where $v,a,b$ are all integers and 
$p$ does not divide $a$ or $b$. The $v\in \ZZ$ is called the \emph{valuation} $v_p(x)$ of $x$; additionally we define $v_p(0):=+\infinity$. 
The $p$-adic norm is defined by $|x|_p:=p^{-v_p(x)}$. 
% (By Ostrowski's theorem these are all the possible norms on $\QQ$.) 
The completion of $\QQ$ under this norm gives rise to the $p$-adic numbers $\QQ_p$. The $p$-adic integers $\ZZ_p$ in $\QQ_p$ are those elements $z\in\QQ_p$ with $|z|_p\leq 1$, i.e.\ those with $v_p(z)\geq 0$. (The ring of $p$-adic integers $\ZZ_p$ should not be confused with the field $\ZZ/p\ZZ$ of integers modulo $p$, although some authors denote this field also by $\ZZ_p$.)
Note that with this norm, the limit
$\lim_{k\rightarrow +\infinity} p^k$ converges to $0$, and hence $1+\lim_{k\rightarrow +\infinity} (p^k-1) = 0$, 
showing that $-1 = p-1 + \sum_{j=2}^{+\infinity} p^j$ in $\QQ_p$. The observation supports the following notation. 

% \subsection{Notation}
Each nonzero $p$-adic number $z\in\QQ_p$ can be described by the formal power series
\begin{align}
z = \sum_{j=v_p(x)}^{+\infinity} z_j p^j \mbox{~with $z_j\in\{0,\dots,p-1\}$ for all $j\geq v_p(z)$ and $z_{v_p(z)}\neq 0$.} 
\end{align}
The sets $\ZZ_p$ and $\QQ_p$ are both uncountable; $\ZZ_p$ is compact, while $\QQ_p$ is only locally compact. Just as the field $\QQ_p$ shares many properties
with $\RR$, the ring $\ZZ_p = \{x\in\QQ_p : |x|_p\leq 1\}$ has several similarities with $\RR/\ZZ \simeq \{x\in\RR : |x|<1\}$.

For each $\QQ_p\ni z = \sum_{j=v}^{\infinity} z_j{p^j}$ its \emph{fractional part} is defined by 
$\{z\} := \sum_{j=v}^{-1}z_j{p^j}$. For all $z\in\QQ_p$ we have that 
$\{z\}$ is a rational number from the set $\{m/p^n : n\in\ZZ^+, m\in\{0,\dots,p^n-1\}\}$; one can also view this value as 
$\{z\} = z\bmod{\ZZ_p}$. We define the function $e:\QQ_p\rightarrow \CC$ by $e(x) = \exp(2\pi\i \{x\})$ such that $|e(x)|=1$ for all $x$. 
Note that, while $\{x+y\}$ does not always equal $\{x\}+\{y\}$, it does hold that $e(x+y) = e(x)e(y)$, making $e$ an additive character of $\QQ_p$ that 
is trivial on $\ZZ_p$. For each $\alpha\in\QQ_p$ the $\QQ_p\rightarrow \CC$ function $x\mapsto e(\alpha x)$ is an additive character on $\QQ_p$ and all characters can be expressed this way, hence $\QQ_p$ is its own Pontryagin dual \cite{FourierNumberField}. For characters over $\ZZ_p$ we have that $e(\alpha x) = e((\alpha+1)x)$ for all $x\in\ZZ_p$, 
which shows that only the fractional part of $\alpha$ matters in this context. Indeed one can show that the dual of $\ZZ_p$ is equivalent to the group $\QQ_p/\ZZ_p$.  

\subsection[Measures on ${\QQ_p}$, quantum states over ${\QQ_p}$, and ${L^2(\QQ_p)}$]{Measures on $\boldsymbol{\QQ_p}$, quantum states over $\boldsymbol{\QQ_p}$, and $\boldsymbol{L^2(\QQ_p)}$}
The standard, normalized Haar measure $\mu$ on $\QQ_p$ is given by $\mu(p^j\ZZ_p) = p^{-j}$ such that $\mu(\ZZ_p)=1$ and, for $z\in\QQ_p$, we have $\mu(z+p^j\ZZ_p) = p^{-j}$ where $z+S = \{z+x : x\in S\}$. Notice that $\mu(\QQ_p) = +\infinity$. From now on, an integral $\int_{x\in\QQ_p}\d x$ is understood to be taken with respect to this measure. The Hilbert space $L^2(\QQ_p)$ has as its elements
\begin{align}
L^2(\QQ_p) & := \{\psi:\QQ_p\rightarrow \CC : \|\psi\|<\infinity\} \\
\intertext{with the $\ell_2$-norm defined by}
\|\psi\| & := \sqrt{\int_{x\in\QQ_p}{\psi(x)\psi^*(x)\d x}}.
\end{align}
A function/quantum state $\psi$ is \emph{normalized} when $\|\psi\|=1$. 
 
\subsection{Fourier transforms in $\boldsymbol{L^2(\QQ_p)}$}
For a function $\psi\in L^2(\QQ_p)$, its \emph{Fourier transform} $\hat{\psi}:\QQ_p\rightarrow \CC$ is defined by 
\begin{align}
\hat{\psi}(y) = \int_{x\in\QQ_p}{\psi(x) e(xy)\d y} \mbox{~for all $y\in\QQ_p$.}
\end{align}

\begin{example}
Fix $r\in\ZZ$ and $z\in\QQ_p$. Define the function $\psi\in L^2(\QQ_p)$ by 
\begin{equation}
\psi(x) = 
\begin{cases}
p^{r/2} & \mbox{if $x\in z+p^r\ZZ_p$}\\
0 & \mbox{otherwise}.
\end{cases}
\end{equation}
Notice that $\psi$ is normalized with $\|\psi\|=1$ and that with the $p$-adic distance between the elements of $\QQ_p$ this indicator function is continuous.  
The Fourier transform of $\psi$ is described by 
\begin{equation}
\hat{\psi}(y) = 
\begin{cases}e(yz)\cdot p^{-r/2} & \mbox{if $y\in p^{-r}\ZZ_p$} \\
0 & \mbox{otherwise.}
\end{cases}
\end{equation}  
As the Fourier transform is a isometry from $L^2(\QQ_p)$ to itself (under the natural inner product given by Haar measure), 
such $\psi$ are the natural $\QQ_p$-equivalents of the Gaussian distribution over $\RR$ with $r$ acting as the scale.
\end{example}

\section{$(p+1)$ Mutually Unbiased Bases in $L^2(\QQ_p)$}
\label{sec:padicMUBs}
Just as in the $L^2(\RR)$ case of Weigert and Wilkinson \cite{MUBsContinuousVariables}, our \MUBs are `generalized' eigenstates, 
meaning that they are not properly normalized. To deal with this issue we will use the standard technique of describing them as a 
limit of functions that \emph{are} elements of  $L^2(\QQ_p)$. The limit that we will rely upon is $\lim_{r\rightarrow+\infinity}p^{-r}\ZZ_p=\QQ_p$ (as sets). (All of this can be made more rigorous by casting it in the framework of `rigged' Hilbert spaces \cite{Ballentine}.)
We allow ourselves the slight abuse of notation where 
\begin{align}
L_2(\QQ_p) \ni \ket{\psi} = \int_{x\in\QQ_p}\psi(x)\ket{x}\d x\mbox{~stands for $\psi:\QQ_p\rightarrow\CC$ with $\psi:x\mapsto \psi(x)$.}
\end{align}

For each $r\in\ZZ$ we define the following sets $V_a^{(r)}\subseteq L^2(\QQ_p)$ indexed by $a\in\QQ_p$:
\begin{align}\label{eq:Vadef}
V_a^{(r)} := \{\ket{v(a,b;r)} : b\in\QQ_p\}
\mbox{~with~}
 \ket{v(a,b;r)} & := \int_{x\in p^{-r}\ZZ_p}e(ax^2+bx)\ket{x} \d x 
\end{align}
The following lemma regarding the norm of quadratic Gauss integrals over $p^{-r}\ZZ_p$ is proven in Lemma~\ref{lem:gausssumQp} in Appendix~\ref{app:gaussintegrals}.
\begin{lemma}\label{lemma:gausssumQp} 
Let $\alpha,\beta\in\QQ_p$ and $r\in \ZZ$, then
\begin{align}
\left|
\int_{x\in p^{-r}\ZZ_p}e(\alpha x^2+\beta x)\d x
\right| & = 
\begin{cases}
p^{v_p(\alpha)/2} & \mbox{if $v_p(\alpha) < 2r$ and $v_p(\alpha)\leq v_p(\alpha)+r$}\\ 
0 & \mbox{if $v_p(\beta) < r$ and $v_p(\alpha)>v_p(\beta)+r$}\\
p^r & \mbox{if $v_p(\alpha)\geq 2r$ and $v_p(\beta)\geq r$}
\end{cases}
\end{align}
\end{lemma}
Additionally we define the set $V_\infinity^{(r)}\subseteq L^2(\QQ_p)$ as
\begin{align}\label{eq:Vinfdef}
V_\infinity^{(r)} := \{\ket{v(\infinity,b;r} : b\in\QQ_p\}
\mbox{~with~}
 \ket{v(\infinity,b;r)} & := p^{r}\int_{x\in p^{r}\ZZ_p}\ket{x-b} \d x 
\end{align}
Note that the $v(\infinity,b;r)$ is the Fourier transform of $v(0,b;r)$ and that the
$V_\infinity$ vectors act like the scaled delta-functions 
$\sqrt{p^r \cdot \delta(x-b)}$ as $r\rightarrow \infinity$. 

In Corollary~\ref{cor:VaMUB} in Appendix~\ref{app:gaussintegrals} it is proven that for all $a,a'\in\QQ_p$ 
and $b,b'\in\QQ_p$ for large enough $r$ we have 
\begin{align}
|\braket{v(a',b';r)}{v(a,b;r)}| & = 
\begin{cases}
p^{v_p(a-a')/2} & \mbox{if $a\neq a'$} \\
0 & \mbox{if $a = a'$ and $b\neq b'$} \\
p^r & \mbox{if $a = a'$ and $b = b'$.}
\end{cases} \\
\intertext{Additionally, with Lemmas~\ref{lem:Vinf} and \ref{lem:deltafunctions} in Appendix~\ref{app:gaussintegrals} we see that with $a'=\infinity$, $a\in\QQ_p\cup\{\infinity\}$ and $b,b'\in\QQ_p$, for large enough $r$ we have the inner product:}
|\braket{v(\infinity,b';r)}{v(a,b;r)}| & = 
\begin{cases}
1 & \mbox{if~$a\neq \infinity$}\\
0 & \mbox{if $a=\infinity$ and $b\neq b'$} \\
p^r & \mbox{if $a=\infinity$ and $b= b'$}. 
\end{cases}
\end{align}

As mentioned before, in the limit $r\rightarrow+\infinity$ we have $p^{-r}\ZZ_p\rightarrow \QQ_p$ as sets. 
Hence with the definitions and results of this section we obtain the main theorem of this article.
\begin{theorem}
With the above definitions, all sets  $V^{(\infinity)}_\infinity$ and $V^{(\infinity)}_a$ with $a\in\QQ_p$  are a basis for $L^2(\QQ_p)$. 
For $a,a'\in\{0,\dots,p-1\}$ it holds that $a-a'=0$ or $v_p(a-a')=0$ and hence we have that $\{V_0^{(\infinity)},\dots,V_{p-1}^{(\infinity)},V_\infinity^{(\infinity)}\}$ makes a set of $p+1$ mutually unbiased bases in $L^2(\QQ_p)$ such that for all $a,a'\in\{0,\dots,p-1,\infinity\}$ we have the unbiased inner product $|\braket{v(a',b';\infinity)}{v(a,b;\infinity)}| = 1$ when $a\neq a'$. 
\end{theorem}
In the $r\rightarrow\infinity$ limit the functions $v(a,b;r) \in V_a^{(r)}$ are described by
\begin{align}
V_a^{(\infinity)}\ni\ket{v(a,b;\infinity)} & = \int_{x\in\QQ_p} e(ax^2+bx)\ket{x}\d x
\end{align}
for all $a,b\in\QQ_p$. The elements of $V_\infinity^{(\infinity)}$ are the Fourier transforms of the $V_0^{(\infinity)}$ functions:
\begin{align}
V_\infinity^{(\infinity)}\ni\ket{v(\infinity,b;\infinity)} & = 
\mathrm{Fourier}_{\QQ_p} \ket{v(0,b;\infinity)}. 
\end{align}
In Appendix~\ref{app:unitaryMUB} we have a description of these $(p+1)$ \MUBs as eigenfunctions of $(p+1)$ families of unitary transformations on $L^2(\QQ_p)$. 

\section{Difference for MUBs between $L^2(\RR)$ and $L^2(\QQ_p)$}
Why is it that we have $p+1$ \MUBs over the $p$-adic numbers $\QQ_p$, while we only know of $3$ \MUBs over the reals $\RR$?

If we want to adapt the $(ax^2+bx)$ construction to the case of quantum states defined over $\RR$, we can use the normalized Gaussian distribution
\begin{align}
\frac{\sqrt{2}}{k\sqrt{\pi}}\int_{x=-\infinity}^{+\infinity} \e^{-2(x/k)^2}\d x & = 1
\end{align}
(which in the limit $k\rightarrow + \infinity$ gives the `uniform distribution over $\RR$'), 
to define the basis states
\begin{align}
\ket{v(a,b;k)} & = \sqrt{\frac{\sqrt{2}}{k\sqrt{\pi}}}\int_{x=-\infinity}^{+\infinity}
\e^{-(x/k)^2}\e^{2\pi\i(ax^2+bx)}\ket{x}\d x. 
\end{align}
For the relevant integral we then get
\begin{align} 
\lim_{k\rightarrow +\infinity }\left|\frac{\sqrt{2}}{\sqrt{\pi}}\int_{x=-\infinity}^{+\infinity} \e^{2\pi\i (\alpha x^2+\beta x)}\e^{-2(x/k)^2}\d x\right| & \propto \frac{1}{\sqrt{|\alpha|}},
\end{align}
which shows that for different bases $B_a$ and $B_{a'}$ we have the dependency
\begin{align}\label{eq:proptoa}
|\langle v(a,b;k) | v(a',b';k)\rangle| & \propto \frac{1}{\sqrt{|a-a'|}}.
\end{align}
% Similarly, the uniform distribution $1/2k$ for $[-k,k]$ as $k\rightarrow +\infinity$ also 
% gives the dependency of Equation~\ref{eq:proptoa} for the relevant integral
% \begin{align}
% \lim_{k\rightarrow +\infinity}\frac{1}{2k}\int_{x=-k}^{+k}\e^{2\pi\i a x^2}\d x & = 
% \frac{1+\i}{4k\sqrt{a}}.
% \end{align}
Note now how the norm of the quadratic Gauss integral in Lemma~\ref{lem:gausssumQp} also has a $1/\sqrt{|a|}$ dependency, 
giving the same kind of dependency of Equation~\ref{eq:proptoa}. 
For $L^2(\QQ_p)$ however, the $|a|$ is the $p$-adic norm with $|a|=p^{-v_p(a)}$, which is much more coarse than the standard $\RR$ norm. 
If $A$ is a set of $a$ coefficients, then $\{B_a : a\in A\}$ defines a set of \MUBs if $|a-a'| = |a''-a'''|$ for all $a\neq a', a''\neq a''' \in A$. 
For the $\RR$-norm the maximum set $A$ with this property is $A=\{0,1\}$, while for the $p$-adic norm we can have $A=\{0,1,\dots,p-1\}$. 

\section{Conclusion and Open Questions}
We have found that there are at least $p+1$ \MUBs if we consider quantum states defined over the $p$-adic numbers $\QQ_p$, while we only know of 
$3$ \MUBs for quantum states defined over the real line $\RR$. An obvious open question is: are these numbers tight, or is it possible to define 
more \MUBs over $\QQ_p$ and/or $\RR$? Upon closer inspection, one can see that the reason why the $(ax^2+bx)$ construction leads to different \textsc{MUB} situations for $\QQ_p$ and $\RR$ stems from the fact 
that their respective norms have different properties. It is another open question if this analysis is only specific to the $(ax^2+bx)$ construction that we used here, or if this in fact \emph{the} reason why we might have different numbers of \MUBs for $L^2(\QQ_p)$ and $L^2(\RR)$.

\subsection*{Acknowledgements}
The research presented in this article was supported in part by an NSF CAREER grant (WvD) 
and by a grant from the Army Research Office (WvD and AR).

\appendix

\section{Quadratic Gauss Sums and Integrals}\label{app:gaussintegrals}
The results in this appendix are not new; see for example Equation~2.11 in \cite{Vladimirov}. The two main reasons to provide these proofs here is to make the currently article self-contained and to give the reader a flavor of how to do calculations in $\QQ_p$.  
\subsection{Quadratic Gauss sums over finite rings $\boldsymbol{\ZZ/p^k\ZZ}$}
The following lemma regarding the norm of quadratic sums over $\ZZ/p^k\ZZ$ will be used to prove
Lemma~\ref{lem:gausssumQp} about the norm of quadratic Gauss integrals over $p$-adic numbers. 
\begin{lemma}\label{lem:gausssumfinitering}
Let $p\neq 2$ be a prime, $k, \ell\in\ZZ^+$ with $k\geq \ell$, $a,b\in\ZZ$ and 
$\omega:= \e^{2\pi\i/p^\ell}$. The quadratic Gauss sum has the following norm 
\begin{align}
\left|{\sum_{x\in\{0,\dots,p^k-1\}}\omega^{ax^2+bx}}\right|
& = 
\begin{cases}
p^{k-\ell/2+v_p(a)/2} & \mbox{if $a\neq 0\bmod{p^\ell}$ and $v_p(a)\leq v_p(b)$}\\ 
0 & \mbox{if $b\neq 0\bmod{p^\ell}$ and $v_p(a)> v_p(b)$} \\
p^k & \mbox{if $a=b=0 \bmod{p^\ell}$.}
\end{cases}
\end{align}
\begin{proof}
Because $\omega^z$ is an additive character over $\ZZ/p^\ell\ZZ$ we can interpret
the values $ax^2+bx$ modulo $p^\ell$. 
Define and note
\begin{align}
g(a,b;k,\ell) := \sum_{x\in\{0,\dots,p^k-1\}} \omega^{ax^2+bx} 
&= p^{k-\ell}\sum_{x\in\ZZ/p^\ell\ZZ} \omega^{ax^2+bx}\\
& = p^{k-\ell}\cdot g(a,b;\ell,\ell). \label{eq:k_equals_l}
\end{align}
% We therefore only have to consider the $k=\ell$ case. 
To ignore the phase of $g$, we look at the norm of $gg^* = |g|^2$:
\begin{align}
|g(a,b;\ell,\ell)|^2 
&= 
\Big(\sum_{x\in\ZZ/p^\ell\ZZ} \omega^{ax^2+bx}\Big) 
\Big(\sum_{y\in\ZZ/p^\ell\ZZ} \omega^{-ax^2-bx}\Big)\\
&= 
\sum_{x,y\in\ZZ/p^\ell\ZZ} \omega^{(x-y)(a(x+y)+b)} \\
\intertext{Because $p\neq 2$ the mapping $(x',y') = (x-y,x+y)$ can be inverted by $(x,y)=((x'+y')/2,(y'-x')/2)$, proving that this mapping is a permutation of $(\ZZ/p^\ell\ZZ)^2$. Hence we can rewrite the summation according to $x\leftarrow (x-y)$ and $y\leftarrow (x+y)$, giving us}
|g(a,b;\ell,\ell)|^2 
&= 
\sum_{x,y\in\ZZ/p^\ell\ZZ} \omega^{x(ay+b)} \\
\intertext{As $\sum_x\omega^{x\delta} = p^\ell$ if $\delta = 0\bmod{p^\ell}$ and $\sum_x\omega^{x\delta} = 0$ if $\delta \neq 0\bmod{p^\ell}$ this simplifies to}
|g(a,b;\ell,\ell)|^2 \label{eq:simplifies_to}
& = p^\ell\cdot |\{y\in\ZZ/p^\ell\ZZ : ay+b=0\bmod{p^\ell}\}|
\end{align}
At this stage we need to calculate how many solutions $y\in\ZZ/p^\ell\ZZ$ there are to the linear equation $ay+b=0\bmod{p^\ell}$, which depends on the values of $a$ and $b$ modulo $p^\ell$. We will have to analyze several cases that can occur depending on whether 
$a$ or $b$ equal $0\bmod{p^\ell}$ and the two valuations $v_p(a)$ and $v_p(b)$.  
If $a\neq 0\bmod{p^\ell}$, we will write $a=\alpha p^{v_p(a)}$ with $p\nmid\alpha$ and $v_p(a)\in\{0,\dots,\ell-1\}$. 
Similarly, if $b\neq 0\bmod{p^\ell}$, we will write $b=\beta p^{v_p(b)}$ with $p\nmid\beta$ and $v_p(b)\in\{0,\dots,\ell-1\}$. 
\begin{itemize}
\item If $a=0\bmod{p^\ell}$ and $b=0\bmod{p^\ell}$, then there are $p^\ell$ solutions $y\in\ZZ/p^\ell\ZZ$. 
\item If $a=0\bmod{p^\ell}$ and $b\neq 0\bmod{p^\ell}$, then there are no solutions.  
\item If $a\neq 0\bmod{p^\ell}$ and $b=0\bmod{p^\ell}$ the equation $\alpha p^{v_p(a)}y=0\bmod{p^\ell}$ can be re-expressed as 
$\alpha y = 0\bmod{p^{\ell-v_p(a)}}$, which has $p^{v_p(a)}$ solutions $y\in\{0,\dots, p^{v_p(a)}-1\}p^{\ell-v_p(a)}$. 
\item If $a,b\neq 0\bmod{p^\ell}$ and $v_p(a)\leq v_p(b)$, there are $p^{v_p(a)}$ solutions the equation $y = (-\beta/\alpha) p^{v_b(b)-v_p(a)} \bmod{p^{\ell-v_p(a)}}$, namely $y\in(-\beta/\alpha)p^{v_p(b)-v_b(a)} + \{0,\dots,p^{v_p(a)}-1\}p^{\ell-v_p(a)}$. 
\item If $a,b\neq 0\bmod{p^\ell}$ and $v_p(a)>v_p(b)$ there are no solutions to the equation $p^{v_p(a)-v_p(b)}y = (-\beta/\alpha) \bmod{p^{\ell-v_p(b)}}$  as $p\nmid(-\beta/\alpha)$. 
\end{itemize}
Summarizing we have
\begin{align}\label{eq:summarizing}
|\{y%\in\ZZ/p^\ell\ZZ 
: ay+b=0\bmod{p^\ell}| & = 
\begin{cases}
p^{v_p(a)} & \mbox{if $a\neq 0\bmod{p^\ell}$ and $v_p(a)\leq v_p(b)$}\\
0 & \mbox{if $b\neq 0\bmod{p^\ell}$ and $v_p(a) > v_p(b)$}\\
p^\ell & \mbox{if $a=b=0\bmod{p^\ell}$.}
\end{cases}
\end{align}
The combination of Equations~\ref{eq:k_equals_l}, \ref{eq:simplifies_to} and \ref{eq:summarizing} proves the lemma. 
\end{proof}
\end{lemma}

\subsection{Quadratic Gauss integrals over $\boldsymbol{p}$-adic numbers}
With Lemma~\ref{lem:gausssumfinitering}, the following result can be proven. 
\begin{lemma}\label{lem:gausssumQp} 
Let $a,b\in\QQ_p$ and $r\in \ZZ$, then
\begin{align}
\left|
\int_{x\in p^{-r}\ZZ_p}e(ax^2+bx)\d x
\right| & = 
\begin{cases}
p^{v_p(a)/2} & \mbox{if $v_p(a) < 2r$ and $v_p(a)\leq v_p(b)+r$} \\
0 & \mbox{if $v_p(b) < r$ and $v_p(a)>v_p(b)+r$}\\
p^r & \mbox{if $v_p(a)\geq 2r$ and $v_p(b)\geq r$.}
\end{cases}
\end{align}
\begin{proof}
Define and rewrite
\begin{align}
g(a,b;p^{-r}\ZZ_p) & := \int_{x\in p^{-r}\ZZ_p}e(ax^2+bx)\d x \\
& = p^r\int_{x\in \ZZ_p}e(a(x/p^r)^2+b(x/p^r))\d x \\
& = p^r\int_{y\in p^k\ZZ_p}\sum_{z\in\{0,\dots,p^k-1\}}e(a((y+z)/p^r)^2+b((y+z)/p^r))\d y \\
& = p^r\int_{y\in p^k\ZZ_p}\sum_{z\in\{0,\dots,p^k-1\}}e\Big(\frac{a(y^2+2yz+z^2)+b(y+z)p^r}{p^{2r}}\Big)\d y \\
\intertext{with $k\in\ZZ^+$. As $y$ is a multiple of $p^k$, for sufficiently large $k$ 
(i.e.\ $k\geq r-v_p(a)/2$, $k\geq 2r-v_p(a)$, $k\geq r-v_p(b)$)
the term $(ay^2+2ayz+byp^r)/p^{2r}$ can be made an element of $\ZZ_p$
for all $z\in\ZZ$, which makes the term irrelevant for the $e$ function. 
This then gives us the finite summation}
& = p^r\int_{y\in p^k\ZZ_p}\sum_{z\in\{0,\dots,p^k-1\}}e\Big(\frac{az^2+bzp^r}{p^{2r}}\Big)\d y \\
& = p^{r-k}\sum_{z\in\{0,\dots,p^k-1\}}e\big(ap^{-2r} z^2+bp^{-r}z\big) \\
\intertext{Pick $\ell\in\ZZ$ such that $\ell\geq 2r-v_p(a)$ and $\ell\geq r-v_p(b)$, making $A=a p^{-2r+\ell}$ and $B=b p^{-r+\ell}$ 
both elements of $\ZZ_p$. We can then rewrite the summation as}
& = p^{r-k}\sum_{z\in\{0,\dots,p^k-1\}}e((A z^2+B z)/p^\ell) \\
& = p^{r-k}\sum_{z\in\{0,\dots,p^k-1\}}\e^{2\pi\i (A z^2+B z)/p^\ell} \\
& = p^{r-k}\cdot g(A,B;k,\ell). 
\end{align}
Assume without loss of generality that $k\geq \ell$.
Using the previous lemma on Gauss sums over $\{0,\dots,p^k-1\}$
and the fact that $v_p(A) = v_p(a)-2r+\ell$ and $v_p(B) = v_p(b)-r+\ell$ we get
\begin{align}
|g(a,b;p^{-r}\ZZ_p)| & = 
p^{r-k}\cdot |g(A,B;k,\ell)|\\
 & = 
\begin{cases}
p^{v_p(a)/2} & \mbox{if $v_p(a) < 2r$ and $v_p(a)\leq v_p(b)+r$} \\
0 & \mbox{if $v_p(b) < r$ and $v_p(a)>v_p(b)+r$}\\
p^r & \mbox{if $v_p(a)\geq 2r$ and $v_p(b)\geq r$.}
\end{cases}
\end{align}
\end{proof}
\end{lemma}

\begin{corollary}\label{cor:VaMUB}
Let $a,b\in\QQ_p$. Define a threshold value $t\in\ZZ\cup\{-\infinity\}$ as follows. 
If $a=0$ and $b=0$ then $t=-\infinity$; 
if $a=0$ and $b\neq 0$ then $t=v_p(b)$; 
if $a\neq 0$ then $t=\max\{v_p(a)/2, v_p(a)-v_p(b)\}$. 
Then, for every $r> t$ we have 
\begin{align}
|g(a,b;p^{-r}\ZZ_p)| & = \label{eq:bigenoughr}
\begin{cases}
p^{v_p(a)/2} & \mbox{if $a\neq 0$}\\
0 & \mbox{if $a=0$ and $b\neq 0$}\\
p^r & \mbox{if $a=b=0$.} 
\end{cases}
\end{align}
\end{corollary}
This last corollary shows that for any $a,b\in\QQ_p$ as $r$ gets `big enough', 
we have the three cases of Equation~\ref{eq:bigenoughr} for the norm of the quadratic Gauss sum. 
Hence for the case $a\neq 0$ or $b\neq 0$ we have the $r\rightarrow +\infinity$ limit
\begin{align}% \label{eq:bigenoughr}
\left|\int_{x\in\QQ_p}e(ax^2+bx) \d x\right| = |g(a,b;\QQ_p)| & = 
\begin{cases}
p^{v_p(a)/2} & \mbox{if $a\neq 0$}\\
0 & \mbox{if $a=0$ and $b\neq 0$.}
\end{cases}
\end{align}

% \begin{corollary}
% On the other hand, for $r=0$, and $v_p(a)<0$ or $a=0$ we have 
% \begin{align}
% |g(a,b;\ZZ_p)| & = 
% \begin{cases}
% 1 & \mbox{if $a=0$ and $v_p(b)\geq 0$} \\
% 0 & \mbox{if $v_p(b)<0$ and $v_p(a)>v_p(b)$} \\
% p^{v_p(a)/2} & \mbox{if $a\neq 0$ and $v_p(a)\leq v_p(b)$.}
% \end{cases}
% \end{align}
% \end{corollary}

The $V_\infinity$ functions of Equation~\ref{eq:Vinfdef} have the same size and orthogonality as the $V_a$ functions as is shown by the following lemma.
\begin{lemma}\label{lem:Vinf}
Use the definitions of Equation \ref{eq:Vinfdef} and let $b,b'\in\QQ_p$. There exists a threshold $t\in\ZZ\cup\{-\infinity\}$ such that for all $r>t$ we have 
\begin{align}
|\braket{v(\infinity,b';r)}{v(\infinity,b;r)}| 
& = \begin{cases}
0 & \mbox{if~$b\neq b'$}\\
p^r &\mbox{if $b=b'$.}
\end{cases}
\end{align}
\begin{proof}
\begin{align}
|\braket{v(\infinity,b';r)}{v(\infinity,b;r)}| & = \left|p^{2r}\int_{x,y\in p^r\ZZ_p}\braket{y-b'}{x-b}\d x\d y\right| \\
& = p^{2r}\mu(\{x\in p^r\ZZ_p : x+b-b'\in p^r\ZZ_p\}) \\
& = \begin{cases}
0 & \mbox{if~$v(b-b') < r$}\\
p^r &\mbox{if $v(b-b')\geq r$}
\end{cases}
\end{align}
Hence we can $t$ be as follows: if $b=b'$, set $t=-\infinity$; if $b\neq b'$, set $t=v(b-b')$. 
\end{proof}
\end{lemma}

The next simple lemma is needed to prove that the $V_\infinity$ functions of the main theorem are mutually unbiased to the $V_a$ functions of Equation~\ref{eq:Vadef}. 
\begin{lemma} \label{lem:deltafunctions}
Let $a,b,c\in\QQ_p$. There exists a threshold $t \in \ZZ$ such that for all $r\geq t$:
% we have
\begin{align}
\left|p^r\int_{x\in c+p^r\ZZ_p}e(a x^2+b x)\d x\right| 
& = 1.
\end{align}
\begin{proof}
\begin{align}
\left|p^r\int_{x\in c+p^r\ZZ_p}e(a x^2+b x)\d x\right| 
& = 
\left|p^r\int_{x\in p^r\ZZ_p}e(a c^2+2 a cx+ax^2+b c+bx)\d x\right| \\
\intertext{For $r\geq\max\{-v_p(2ac+b),-v_p(a)/2\}$ we have that $2 a c x+ax^2+bx \in\ZZ_p$ for all $x\in p^r\ZZ_p$, 
which simplifies the integral to }
& = 
\left|p^r\int_{x\in p^r\ZZ_p}e(a c^2+b c)\d x\right| = 1.
\end{align}
\end{proof}
\end{lemma}

\section{A Unitary Description of the $(p+1)$ MUBs in $L^2(\QQ_p)$} \label{app:unitaryMUB}
As was mentioned in Section~\ref{sec:L2RMUBs}, the three \MUBs in $L^2(\RR)$ can be described as the eigenstates of three Hermitian operators that are combinations of the momentum operator $\hat{p}$ and the position 
operator $\hat{q}$. When trying to mimic this approach for the $(p+1)$ \MUBs in $L^2(\QQ_p)$ one quickly realizes that some complications arise due to the difference between $\QQ_p$ and $\RR$. 
A very concrete example of this complication is the fact that the position operator $\hat{q}:\ket{x}\mapsto x\ket{x}$
does not make sense in this setting as it tries to assign the $\QQ_p$ valued position value $x$ as an amplitude.  
As a way out, we will we will omit the Hamiltonian description and instead capture the \MUBs in terms of the eigenfunctions of $(p+1)$ families of unitary transformations on $L^2(\QQ_p)$. 

Define the two sets of unitary operators $X_c$ and $Z_d: L^2(\QQ_p)\rightarrow L^2(\QQ_p)$ 
with the parameters $c,d\in\QQ_p$ by
\begin{align}
X_c:\ket{y} \mapsto \ket{y+c} \mbox{~and~}Z_d:\ket{y} \mapsto e(yd)\ket{y}~\mbox{~for all $y\in\QQ_p$.}
\end{align}
(As $X_c Z_d:\ket{y}\mapsto e(yd)\ket{y+c}$ and $Z_d X_c:\ket{y}\mapsto e(yd+cd)\ket{y+c}$, we have the expected relation $Z_dX_c = e(cd)X_cZ_d$.)  

For $a,b,c\in\QQ_p$ take $d=2ac$ and observe that 
\begin{align}
X_cZ_{2ac}:\ket{v(a,b;\infinity)} & := X_cZ_{2ac}:\int_{x\in\QQ_p}e(ax^2+bx)\ket{x}\d x \\
& \mapsto 
\int_{x\in\QQ_p}e(ax^2+bx+2acx)\ket{x+c}\d x \\
& = 
\int_{x\in\QQ_p}e(a(x-c)^2+b(x-c)+2ac(x-c))\ket{x}\d x \\
& = e(-bc-ac^2)\int_{x\in\QQ_p}e(ax^2+bx)\d x \\
& = e(-bc-ac^2)\ket{v(a,b;\infinity)}, 
\end{align}
which shows that the $v(a,b;\infinity)$ in $V^{(\infinity)}_a$ are the eigenfunctions of $X_{c}Z_{2ac}$ (for all $c\in\QQ_p$). Additionally, informally speaking, the (generalized) eigenfunctions of $Z_d$ are those proportional to $\ket{y}$, 
which are exactly the elements of $V_\infinity^{(\infinity)}$.   

Define the following unitary operator $P_d:L^2(\QQ_p)\rightarrow L^2(\QQ_p)$ for all $d\in\QQ_p$ by
\begin{align}
P_d:\ket{x} & \mapsto e(dx^2)\ket{x}\mbox{~for all $x\in\QQ_p$.}
\end{align} 
We immediately have $P_d:\ket{v(a,b;\infinity)} \mapsto \ket{v(a+d,b;\infinity)}$. 
We also have, again informally speaking, $\mathrm{Fourier}_{\QQ_p}:\ket{v(0,b;\infinity)}
\mapsto \ket{v(\infinity, b;\infinity)}$. Hence all $V^{(\infinity)}_a$ bases can be obtained
through a unitary transformation from the $V_{0}^{(\infinity)}$ basis. 
As $V_0^{(\infinity)} = \{\ket{v(0,b;\infinity)}:b\in\QQ_p\}$ is exactly the set of 
characters $x\mapsto e(bx)$ of $\QQ_p$, it is the `Fourier basis' of $L^2(\QQ_p)$. 
By the just described unitary relation between the different bases, we thus have that $V_a^{(\infinity)}$
is a basis of $L^2(\QQ_p)$ for all $a\in\QQ_p\cup\{\infinity\}$. 

\begin{bibdiv}
\begin{biblist}[\normalsize]
% \bib{Applebyetal}{report}{
%  author={D.M.\ Appleby}, 
%  author={Hoan Bui Dang}, 
%  author={Christopher A.\ Fuchs}, 
%  title={Symmetric Informationally-Complete Quantum States as Analogues to   
%  Orthonormal Bases and Minimum-Uncertainty States},
%  date={2007}, 
%  eprint={arXiv:0707.2071v2}
%  }

\bib{Alltop}{article}{
 author={W.O.\ Alltop}, 
 title ={Complex sequences with low periodic correlations}, 
 journal = {IEEE Transactions on Information Theory}, 
 volume = {26}, 
 number ={3}, 
 pages = {350--354}, 
 date = {1980}
}

\bib{Ballentine}{book}{
author={Leslie E.\ Ballentine}, 
title={Quantum Mechanics: a modern approach}, 
publisher={World Scientific Publishing}, 
date={1998}
}

\bib{GaussSums}{book}{
author={Bruce C.\ Berndt}, 
author={Ronald J.\ Evans}, 
author={Kenneth S.\ Williams}, 
title={Gauss and Jacobi Sums}, 
publisher={Wiley-Interscience}, 
series={Canadian Mathematical Society Series of Monographs and Advanced Texts}, 
volume={21}, 
date={1998}
}

\bib{Blume-Kohout}{article}{
 author = {Robin Blume-Kohout}, 
 title = {MUBs in infinite dimensions, the problematic analogy between $L^2(\RR)$ and $\CC^d$}, 
 eprint = {http://pirsa.org/08100072/}, 
 conference={
   title={Seeking SICs: A Workshop on Quantum Frames and Designs}, 
   date={2008}, 
   address={Waterloo, Ontario, Canada}
   },
  organization={Perimiter Institute}
}

\bib{Cerfetal}{article}{
 author={Nicholas J.\ Cerf}, 
 author={Mohammed Bourennane}, 
 author={Anders Karlsson}, 
 author={Nicolas Gisin}, 
 title={Security of Quantum Key Distribution Using $d$-Level Systems}, 
 journal={Physical Review Letters}, 
 volume={88}, 
 number={12}, 
 pages={127902}, 
 date={2002}
} 

\bib{Cormicketal}{article}{
 author={Cecilia Cormick}, 
 author={Ernesto F.\ Galv{\~a}o}, 
 author={Daniel Gottesman}, 
 author={Juan Pablo Paz},
 author={Arthur O.\ Pittenger}, 
 title={Classicality in Discrete Wigner Functions}, 
 journal={Physical Review A}, 
 volume={73}, 
 number={1}, 
 pages={012301}, 
 date={2006}
}

\bib{Gouvea}{book}{
author={Fernando Q.\ Gouv{\^e}a}, 
title={$p$-adic Numbers: an introduction},
edition={Second Edition}, 
series={Universitext}, 
publisher={Springer}, 
date={2000}
}

\bib{ConstructionsofMUBs}{article}{
 author={Andreas Klappenecker}, 
 author={Martin R{\"o}tteler}, 
 title={Constructions of Mutually Unbiased Bases}, 
 conference = {
    title={Finite Fields and Applications, 7th International Conference, Fq7},
    date={2003},   
  },  
  book={
    series={Lecture Notes in Computer Science}, 
    volume={2948}, 
    publisher={Springer}
    },
 pages={137--144}, 
 eprint={arXiv:quant-ph/0309120v1}
}

\bib{Kozyrev}{article}{
 title={Wavelet theory as $p$-adic spectral analysis}, 
 author = {S.V.\ Kozyrev}, 
 journal = {Izvestiya: Mathematics}, 
 volume = {66}, 
 date = {2002}, 
 number = {2}, 
 pages = {367--376}
}

\bib{FourierNumberField}{book}{
 title = {Fourier analysis on number fields}, 
 author ={Dinakar Ramakrishnan}, 
 author = {Robert J.\ Valenza},
 series={Graduate Texts in Mathematics}, 
 volume = {186},  
 publisher ={Springer}, 
 address = {New York}, 
 date = {1999}
}

\bib{QMonpadicfields}{article}{
author={Ph.\ Ruelle}, 
author={E.\ Thiran}, 
author={D.\ Verstegen}, 
author={J.\ Weyers}, 
title={Quantum mechanics on $p$-adic fields}, 
journal={Journal of Mathematical Physics}, 
volume={30}, 
number={12},
pages={2854--2874}, 
date={1989}
}


\bib{Vladimirov}{article}{
author={V.S.\ Vladimirov}, 
author={I.V. Volovich}, 
title={$p$-Adic Quantum Mechanics}, 
journal = {Communications in Mathematical Physics}, 
volume={123}, 
pages={659--676}, 
date={1989}
}

\bib{MUBsContinuousVariables}{article}{
author = {Stefan Weigert},
author = {Michael Wilkinson},
title = {Mutually Unbiased Bases for Continuous Variables}, 
journal={Physical Review A}, 
volume={78}, 
number={2}, 
pages={020303(R)}, 
date = {2008}, 
eprint = {arXiv:0802.03942v2}
}

\bib{WocjanBeth}{article}{
author = {Pawel Wocjan}, 
 author={Thomas Beth},
 title={New Construction of Mutually Unbiased Bases in Square Dimensions}, 
journal = {Quantum Information \& Computation}, 
volume ={5}, 
number={2}, 
pages = {93--101}, 
date = {2005}, 
 eprint={arXiv:quant-ph/0407081}
 }

\bib{WoottersFields}{article}{
 author={William K.\ Wootters}, 
 author={Brian D.\ Fields},
 title={Optimal state-determination by mutually unbiased measurements}, 
 pages={363--381}, 
 volume={191}, 
 number={2}, 
 date={1989}, 
 journal={Annals of Physics}
}
\end{biblist}
\end{bibdiv}
\end{document}